\documentclass[a4paper,11pt]{article}
\pdfoutput=1 

\usepackage{jinstpub} 
\usepackage{array}

\title{\boldmath Optimizing Spin Dressing Sensitivity for the  nEDMSF Experiment}
 
\author[a,1]{V. Cianciolo\note{This manuscript has been authored by UT-Battelle, LLC, under contract DE-AC05-00OR22725 
with the US  Department of Energy (DOE). The US government retains and the publisher, by 
accepting the article for publication, acknowledges that the US government retains a 
nonexclusive, paid-up, irrevocable, worldwide license to publish or reproduce the published 
form of this manuscript, or allow others to do so, for US government purposes. DOE will 
provide public access to these results of federally sponsored research in accordance with the 
DOE Public Access Plan 
\href{http://energy.gov/downloads/doe-public-access-plan}{(http://energy.gov/downloads/doe-public-access-plan).}}}

\affiliation[a]{Oak Ridge National Laboratory,\\Oak Ridge, Tennessee, USA}

\emailAdd{cianciolotv@ornl.gov}

\abstract{

nEDMSF aims to measure the neutron electric dipole moment ($d_n$) with unprecedented precision.
In this paper we explore the experiment's sensitivity when operating with  
an implementation of the critical dressing method in which the angle between the neutron and
Helium-3 spins ($\phi_{3n}$) is subjected to a square modulation by an amount $\phi_d$ (the "dressing angle").
Several parameters can be tuned to optimize sensitivity. 
We find roughly 10\% improvement over a previous estimate, resulting primarily from the addition 
of a waiting period between the $\pi/2$ pulse that initiates $d_n$-driven $\phi_{3n}$ growth and 
the start of $\phi_{3n}$ modulation. We find negligible further improvement by allowing 
$\phi_d$ to vary continuously over the course of a run, and no degradation resulting from 
the addition of an {\it in situ} background measurement into each $\phi_{3n}$ modulation 
sequence. A complete simulation confirms a 300 live-day sensitivity of 
$\sigma = 1.45\times10^{-28}\,e \cdot \rm{cm}$. At this level of sensitivity, 
$\sigma_{\phi_{3n0}} = 1\,{\rm mrad}$ precision on the initial $n/^3{\rm He}$ angle difference
is not negligible. 

} 


\arxivnumber{2410.19033} 

\begin{document}
\maketitle
\flushbottom

\section{Introduction}
\label{sec:intro}

The nEDMSF initiative will develop a new cryogenic neutron electric dipole experiment based on the 
Golub-Lamoreaux proposal~\cite{gollam} and drawing on the legacy nEDM@SNS experiment~\cite{edmjinst}.  
This experiment will measure the neutron electric dipole 
moment ($d_n$) with unprecedented precision to shed light on the source of charge-parity 
violation responsible for the generation of matter in the Universe~\cite{sakharov, psw}. Ultracold 
neutrons (UCN) are created and stored in a pair of material traps ("measurement cells") and 
exposed to highly uniform collinear magnetic and electric fields. The two measurement cells nominally 
have the same magnetic field and opposite electric field. If $d_n$ is non-zero, the neutron precession 
frequency will be altered by an amount proportional to $\vec{E}\cdot\hat{B}$. A small 
concentration (0.1-1 ppb) of polarized $^3$He injected into the measurement cells serves 
as both spin analyzer and comagnetometer.

nEDMSF uses two distinct operating methods ("free precession" and "critical dressing") to 
measure the neutron precession frequency. In both methods, the frequency is determined  
by measuring scintillation light~\cite{lc1, lc2} created in 
$n+^3\!\!{\rm He} \rightarrow p+t+764\,\rm{keV}$ capture events.  
The rate of such events is strongly dependent on the 
angle between the UCN and $^3$He spins ($\phi_{3n}$).
This paper considers optimization of the achievable sensitivity%
\footnote{Quoted here as the $1\sigma$ error bar following 300 live days of operation.}
of the critical dressing method, implemented with square-wave modulation of $\phi_{3n}$.

In Section~\ref{sec:sqmcd} we provide details on the square-wave-modulated critical dressing method.
In Section~\ref{sec:params} we list the experiment parameters involved in the sensitivity calculation
and delineate between fixed ("performance") parameters and tunable ("operating") parameters.
In Section~\ref{sec:opt} we derive a formula for the experimental sensitivity and present optimal
values for the operating parameters. 
In Section~\ref{sec:conc} we present conclusions.

\section{Square-Wave-Modulated Critical Dressing}\label{sec:sqmcd}

The gyromagnetic ratio of particles in a static magnetic field ($B_0$) simultaneously exposed to a 
perpendicular oscillating magnetic field (magnitude $B_d$, frequency $\omega_d$) will effectively be modified, a technique referred to 
as "spin dressing"~\cite{CT}. Golub and Lamoreaux~\cite{gollam} realized that 
with a suitable choice of $B_d$ and $\omega_d$
the effective gyromagnetic ratios of neutrons and $^3$He are equalized.
The criteria is $\gamma_n'\equiv\gamma_nJ_0\left(\frac{\gamma_nB_d}{\omega_d}\right)=
               \gamma_3J_0\left(\frac{\gamma_3B_d}{\omega_d}\right)\equiv\gamma_3'$,
which is satisfied when
$J_d\equiv \frac{4\pi}{h}J_0(\frac{\gamma_nB_d}{\omega_d})\approx \frac{4\pi}{h}0.677$.  
Here, $J_0$ is the zeroth-order Bessel function of the first kind, $\gamma_n$ and $\gamma_3$ 
               are the inherent neutron and $^3$He gyromagnetic ratios, and $\gamma_n'=\gamma_3'$ is the 
               effective gyromagnetic ratio of both species while exposed to the dressing field.
              
Because the $n+^3$He interaction is spin-dependent, UCN moving in a gas of polarized $^3$He
are subject to a pseudomagnetic field ($\vec{B}_p$). 
When "critically dressed", the two species have nearly the same 
precession frequency about the $B_0$ direction (exactly the same if $d_n=0$).
Thus, the angle $\phi_{3n}$ is essentially constant and 
$\vec{B}_p$ will cause the UCN spins to precess around the $^3$He spins~\cite{dwpseudob}, reducing experimental sensitivity. 

In~\cite{gollam}, a sinusoidal modulation of $\phi_{3n}$ was proposed and it was shown that by 
continuously adjusting $B_d/\omega_d$ to eliminate the first harmonic 
term in the scintillation rate (arising from a non-zero $d_n$) the effects of $\vec{B}_p$ could be effectively suppressed.
Alternatively, the dressing field can be modulated such that $\phi_{3n}$ is toggled between different states
\begin{eqnarray}
  ``\!+\!" &:& \phi_{3n} = \phi_{3n0} -\widehat{EB}\tilde{\omega} t  + \phi_{d}\\
  ``\!-\!" &:& \phi_{3n} =  \phi_{3n0} -\widehat{EB}\tilde{\omega}t - \phi_{d}
\end{eqnarray}
\noindent rapidly enough that the precession about the pseudomagnetic field 
can be ignored~\cite{edmjinst}. Here we have used $\widehat{EB}\equiv\hat{E}\cdot\hat{B}$ and refer to $\phi_{d}$ as the "dressing angle".
\begin{equation}\label{eq:domega}
  \tilde{\omega} = J_d|E|\tilde{d}_n
\end{equation}
\noindent is the frequency shift resulting from the combination of the true $d_n$, a false $d_n$ 
if one has been injected into the datastream to blind the analysis~\cite{blind}, the gravitational 
offset between the UCN and the $^3$He comagnetometer~\cite{dwgrav}, the combination of perturbative magnetic fields arising from 
residual magnetic field gradients and the applied electric field~\cite{dwgrav,dweb1,dweb2,dweb3,dweb4,dweb6,dweb7},
and any static offset between $\gamma_n'$ and $\gamma_3'$.
We refer to this approach as "square-wave-modulated critical dressing".

The number of events in the different dressing modulation states is given by~\cite{n3He1,n3He2,n3He3}
\begin{equation}
  N_{\pm} = \left(\frac{N\epsilon_3}{\tau_3}\right)\left(1-P\cos(\phi_{3n0} -\widehat{EB}\tilde{\omega}t  \pm \phi_{d})
                    \right)e^{-\Gamma t}+
                    \left(\frac{N\epsilon_{\beta}}{\tau_{\beta}}\right)e^{-\Gamma t}+\dot{N}_B
\end{equation}
\noindent 
where $N$ is the number of neutrons when the $\pi/2$ pulse rotates the UCN and $^3$He spins into the plane perpendicular to
the field direction to start $d_n$-driven $\phi_{3n}$ accumulation, 
$P=P_3(0)P_n(0)e^{-\Gamma_dt}$ is the time-dependent product of the polarization of the two species, 
$\Gamma_d$ is the combined depolarization time of the two species,
$\Gamma=\frac{1}{\tau_{\beta}} + \frac{1}{\tau_u} + \frac{1}{\tau_w} + \frac{1-P\cos(\phi_d)}{\tau_{3}}$ 
is the total neutron disappearance rate,
$\tau_{\beta}$ is the neutron $\beta$ decay lifetime, 
$\tau_3$ is the spin-averaged $n+^3$He capture time,
$\tau_u$ is the UCN thermal upscattering time constant,
$\tau_w$ is the time constant for UCN to be lost to scattering off the measurement cell walls,
$\epsilon_3$ is the probability to correctly identify a capture event as a signal event, 
$\epsilon_{\beta}$ is the probability to incorrectly identify a $\beta$ decay event as a signal event, and
$\dot{N}_B$ is the rate of non-UCN background events.

Previously%
\footnote{See Equation 2.25 in~\cite{edmjinst} - note the opposite $\phi_{3n}$ sign convention},
it was shown that the difference between $N_+$ and $N_-$ can be formulated as an 
asymmetry that grows linearly in time with a slope proportional to $\tilde{\omega}$ (and thus $\tilde{d}_n$):
\begin{eqnarray}\label{eq:asym1}
  A(t) &=& \frac{N_--N_+}{N_-+N_+} \nonumber \\
        &=&  \frac
                 {\left(\frac{N\epsilon_3}{\tau_3}\right) P\sin\phi_d e^{-\Gamma t} \left(\widehat{EB}\tilde{\omega}t + \phi_{3n0}\right)}
                 {\left(\frac{N\epsilon_3}{\tau_3}\right)\left(1-P\cos\phi_d\right)e^{-\Gamma t}+
                  \left(\frac{N\epsilon_{\beta}}{\tau_{\beta}}\right)e^{-\Gamma t}+\dot{N}_B}
\end{eqnarray}
\noindent Note that this derivation assumes $\phi_{3n0}=0$ and uses the small-angle approximation,  
$\cos(-\widehat{EB}\tilde{\omega}t  \pm \phi_{d}) \approx \cos\phi_d\pm\widehat{EB}\tilde{\omega}t\sin\phi_d$.
In this paper we explore three questions:
\begin{enumerate}

  \item The prospect of fitting Equation~\ref{eq:asym1} is unpleasant because of the many variables beyond 
  $d_n$ that need to be fit. One can imagine obtaining values for these parameters with a set of auxiliary measurements,  
  but that requires additional run time and it is difficult to ensure that conditions in the background and signal runs are 
  sufficiently identical. One alternative is to introduce a ``$0$'' state in each dressing modulation cycle in which
  $\phi_{3n} = \phi_{3n0}  - \tilde{\omega}t $ to provide an {\it in situ} determination of the 
  contribution to the observed event rate from $\beta$ decays, cosmic rays, unpolarized $n+^3$He 
  capture and Compton scattering by activation and ambient $\gamma$'s.
  But does this negatively impact sensitivity?

  \item In a previous sensitivity analysis, square-wave modulation of the dressing angle was assumed 
  to start immediately after the $\pi/2$ pulse that rotates the spins into the plane perpendicular to $B_0$
  to start the $d_n$-driven phase accumulation. The analyzing power of a given modulation 
  cycle is proportional to the phase accumulation, and therefore proportional to the time since the $\pi/2$ 
  pulse. As a result, early cycles have poor analyzing power, but they still consume neutrons. 
  Can sensitivity be improved by waiting for some time ($t_w^d$) after the $\pi/2$ pulse before starting the modulation?

  \item Assuming the answer to 2) is "yes", is there further improvement that can be obtained by gradually increasing 
  $\phi_d$, rather than a step-wise change at $t_w^d$?

\end{enumerate}

\section{Performance Parameters and Operating Parameters}\label{sec:params}

Tables~\ref{tab:pparam} and~\ref{tab:oparam} list parameters used to compute the sensitivity. These are divided 
into "performance parameters" (Table~\ref{tab:pparam}) and "operating parameters" (Table~\ref{tab:oparam}).
Performance parameters are fixed at the values the collaboration expects to ultimately achieve. Operating parameter 
values will be chosen to optimize the sensitivity.

\begin{table}
  \begin{center}
    \begin{minipage}[p]{13cm}
      \begin{tabular}{|l|l|l|} \hline
Parameter & Definition & Value \\ \hline\hline
$\dot{N}_{UCN}$ & UCN production rate & 848 UCN/s%
\footnote{See~\cite{CABaker}, Equation 1. We assume the measured 8.9\AA~fluence out of the Spallation Neutron Source (SNS) 
Fundamental Neutron Physics Beamline (FnPB) cold beamline ($3.4\times 10^7 n/{\rm s}/{\rm cm}^2/{\rm MW}$~\cite{FnPB}) 
scaled up to an SNS power of 2\,MW, neutron transmission efficiency of 11.3\% from the end of the FnPB guide into an individual cell, 
an areal density compression factor of $120\,{\rm cm}^2/70\,{\rm cm}^2$, 160\,neV Fermi potential for the measurement cells' 
deuterated tetraphenylbutadiene + deuterated polystyrene surface and a 3,000\,cc measurement cell volume.}  \\ \hline
$\tau_{\beta}$ & Neutron $\beta$ decay time constant & 879.4\,s~\cite{PDG2020}\\ \hline
$\tau_w$ & Neutron wall loss time constant & 1,974\,s%
\footnote{To simplify the calculation, a single exponential form is assumed for wall loss. The assumed time constant corresponds 
to an 80\,neV UCN in a measurement cell with a Fermi potential of $160-18.5=141.5$\,neV relative to superfluid $^4$He and a 
wall loss constant of $f=8\times10^{-6}$.} \\ \hline
$\tau_u$ & UCN upscattering time constant & 31,321\,s%
\footnote{Assuming an operating temperature $T=0.44$\,K and $\tau_u = 100/T^7$ \cite{upscat1,upscat2}}\\ \hline
$|E|$ & Electric field & 75 kV/cm \\ \hline
$t_d$ & Dead time between measurement cycles & 337\,s \\ \hline
$P_3$ & $^3$He initial polarization & 0.98 \\ \hline
$P_n$ & UCN initial polarization & 0.98 \\ \hline
$\Gamma_d$ & Combined $^3$He and UCN depolarization time & 10,000\,s%
\footnote{Consistent with $10^{-7}$/bounce wall depolarization probability~\cite{LPB1, LPB2} and 
$\partial B_x/d z = 0.8$\,nT/m (the holding field gradient in the long direction of the meaurement cell~\cite{T2}). } \\ \hline
$\epsilon_3$ &  $n+^3$He capture event detection efficiency & 0.93~\cite{lc1} \\ \hline
$\epsilon_{\beta}$ & Neutron $\beta$ decay mis-identification probability & 0.5~\cite{lc1}\ \\ \hline
$\dot{N}_B$ & Non-UCN background rate& 5 Hz%
\footnote{Assumed to be time-independent.} \\ \hline
$\dot{\phi}_d$ & Dressing angle transition rate & 8 rad/s \\ \hline 
$t_l$ & Assumed live time & 300 days \\ \hline
\hline
\end{tabular} 
\caption{nEDMSF performance parameters, fixed at their ultimate values.}
\label{tab:pparam}
\end{minipage}
\end{center}
\end{table}

\begin{table}
\begin{center}
\begin{tabular} {|l|l|c|} \hline
Parameter & Definition & Optimal Value \\ \hline\hline
$t_f$ & Cold neutron fill time  & 875\,s \\ \hline
$t_m$ & Measurement time & 1,085\,s \\ \hline
$\tau_3$ & $^3$He/UCN absorption time constant & 75\,s \\ \hline 
$t_w^d$ & Time between $\pi/2$ pulse and start of dressing modulation & 295\,s \\ \hline
$\phi_d$ & Dressing angle & 0.52\,rad\\ \hline \hline
\end{tabular}
\caption{nEDMSF operating parameters (can be tuned to optimize sensitivity) and optimal 
values assuming performance parameters listed in Table~\ref{tab:pparam}.}
\label{tab:oparam}
\end{center}
\end{table}

\section{Operating Parameter Optimization}\label{sec:opt}

Previously~\cite{edmjinst} it was shown that Equation~\ref{eq:asym1} leads to a formula for
the inverse squared sensitivity of a single run (assuming negligible 
uncertainty on $\phi_{3n0}$, discussed more later):
\begin{equation}\label{eq:uncer}
  \frac{1}{\sigma_{\tilde{d}_n}^2} = 2\left(J_d|E|\right)^2 
  \sum{
    \frac{\left(\left(\frac{N\epsilon_3}{\tau_3}\right)P\sin\phi_de^{-\Gamma t}t\right)^2\Delta t }
          {\left(\frac{N\epsilon_3}{\tau_3}\right)\left(1-P\cos\phi_d\right)e^{-\Gamma t}+
            \left(\frac{N\epsilon_{\beta}}{\tau_{\beta}}\right)e^{-\Gamma t}+\dot{N}_B}
  }
\end{equation}
\noindent where the sum is over modulation cycles in the run. 

The total number of runs is given by:
\begin{equation}
 R=\frac{2t_l}{t_f+t_m+t_d}
\end{equation}

\noindent where $t_l$ is the total experiment live time, $t_f$ is the time spent filling the mreasurement cells with UCN, 
$t_m$ is the time spent making the measurement, $t_d$ is the additional dead time between measurements,
and the ``$2$'' in the numerator accounts for the experiment's two measurement cells.
Scaling $\sigma_{\tilde{d}_n}$ by the number of runs, gives the overall uncertainty:
\begin{eqnarray}\label{eq:nruns}
  \sigma_{d_n} &=& \left( \frac{R}{\sigma^2_{\tilde{d}_n}} \right)^{-1/2} \\
                &=& \frac{1}{2 J_d|E|}
                        \left(
                          \frac{t_l}{t_f+t_m+t_d}
                          \sum{
                             \frac {\left(\left(\frac{N\epsilon_3}{\tau_3}\right)P\sin\phi_de^{-\Gamma t}t\right)^2\Delta t }
                                     {\left(\frac{N\epsilon_3}{\tau_3}\right)\left(1-P\cos\phi_de^{-\Gamma t}\right)+
                                       \left(\frac{N\epsilon_{\beta}}{\tau_{\beta}}e^{-\Gamma t}\right)+\dot{N}_B}
                          } \right)^{-1/2} \nonumber
\end{eqnarray}

We can incorporate a measurement of the number of counts in the ``$0$'' state into the asymmetry:
\begin{equation}
  A(t) = \frac{N_- - N_+}{N_- + N_+ - 2N_0\left(\frac{\Delta t_{\pm}}{\Delta t_0}\right)}
\end{equation}
\noindent where
\begin{equation}
  N_0 = \left(\frac{N\epsilon_3}{\tau_3}\right)\left(1-P\cos(\phi_{3n0} - \widehat{EB}\tilde{\omega} t)
                    \right)e^{-\Gamma t}+
                    \left(\frac{N\epsilon_{\beta}}{\tau_{\beta}}\right)e^{-\Gamma t}+\dot{N}_B
\end{equation}
\noindent and $\Delta t_{\pm}$ and $\Delta t_0$ are the time spent measuring in the 
``$+$'',``$-$'' and ``$0$'' states.
From this we obtain a much simpler relationship between the  asymmetry and $\tilde{\omega}$,
\begin{equation}\label{eq:tofit}
      A(t) = \tan(\phi_{3n0}-\widehat{EB}\tilde{\omega}t)\frac{\sin\phi_d}{\cos\phi_d-1} 
\end{equation}
\noindent This formulation is exact (not relying on the small-angle approximation) 
and seamlessly incorporates non-zero $\phi_{3n0}$. 

If we again ignore uncertainty on $\phi_{3n0}$, we 
find that Equation~\ref{eq:nruns} is still a correct expression for the experimental uncertainty, 
as long as we account for time spent in the ``$0$'' state ($\Delta t \rightarrow \Delta t_{\pm}$). 
Since we wish to allow $\phi_d$, $\Delta t_{\pm}$, and $\Delta t_0$ to be different for different modulation cycles, we
rewrite Equation~\ref{eq:nruns} in terms of $N'=Ne^{-\Gamma t}$, 
the number of events at the mid-point of the modulation cycle:

\begin{equation}\label{eq:final}
\sigma_{d_n} = \frac{1}{2 J_d|E|}
\left(
\frac{t_l}{t_f+t_m+t_d}
\sum{
\frac {\left(\left(\frac{N'\epsilon_3}{\tau_3}\right)P\sin\phi_d t\right)^2\Delta t_{\pm} }
       {\left(\frac{N'\epsilon_3}{\tau_3}\right)\left(1-P\cos\phi_d\right)+
          \left(\frac{N'\epsilon_{\beta}}{\tau_{\beta}}\right)+\dot{N}_B}
} \right)^{-1/2}
\end{equation}

\noindent Now we compute $t$, $N'$ and $P$ for each modulation cycle. The duration of the $i^{\rm th}$ cycle is given by:
\begin{equation}
\Delta t_i =2\Delta t_{\pm}^i + \Delta t_0^i + 2\frac{\phi_d^i}{\dot{\phi}_d} 
\end{equation}
\noindent where the first term accounts for time spent in the ``+" and ``-" states, the second term accounts for 
 time spent in the ``0" state,  and 
the last term accounts for state transition time. From this we obtain the cycle's temporal midpoint:

\begin{eqnarray}
t^i &=& t_w^d \hspace*{0.53in}\mbox{: $i=0$} \nonumber\\
&=& t^{i-1} + \Delta t_i  \hspace*{0.1in} \mbox{: otherwise}
\end{eqnarray}

\noindent polarization:
\begin{equation}
  P^i =P_3(0)P_n(0)e^{-\Gamma_dt^i}
\end{equation}

\noindent and number of neutrons: 

\begin{eqnarray}
N^i &=& \frac{\dot{N}_{UCN}}{\Gamma}(1-e^{-\Gamma t_f})\hspace*{0.1in}\mbox{: $i=0$} \nonumber\\
      &=& N^{i-1}e^{-X^i}  \hspace*{0.67in} \mbox{: otherwise}
\end{eqnarray}

\noindent where

\begin{equation}
X^i = \Gamma_{nc}\Delta t^i + 2\Gamma_{\pm}^i\Delta t_{\pm}^i  + \Gamma_0^i\Delta t_0^i
                                           + 2\Gamma_{\Delta\phi}^i\frac{\phi_d^i}{\dot{\phi}_d}
\end{equation}

\noindent and the four terms on the right-hand side are losses due to 
1) non-capture processes over the entire cycle, 
2) capture losses during the ``+" and ``-" states, 
3) capture losses during the ``0" state, and 
4) capture losses during state transitions. Here we have used

\begin{eqnarray}
\Gamma_{nc} &=& \frac{1}{\tau_{\beta}} + \frac{1}{\tau_u} + \frac{1}{\tau_w} \\
\Gamma_{\pm}^i &=& \frac{1-P^i\cos\phi_d^i}{\tau_3} \\
\Gamma_{\Delta\phi}^i &=&  \frac{1-P^i\cos(\phi_d^i/2)}{\tau_3}\\
\Gamma_0^i &=&  \frac{1-P^i}{\tau_3}
\end{eqnarray}

We define "initial operating parameter values" to be those used in~\cite{edmjinst} ($t_f=1000$\,s, $t_m=1000$\,s, 
$\tau_3=100$\,s, $\phi_d=0.48$\,rad, $t_w^d=0$\,s, $\Delta t_0=0$\,s, $\dot{\phi}_d=\infty$).
Using these values we find the same statistical uncertainty on $d_n$, $\sigma_{d_n}=1.6\times 10^{-28} e\cdot{\rm cm}$
that was found in~\cite{edmjinst}.

\begin{figure}
\includegraphics[width=13cm]{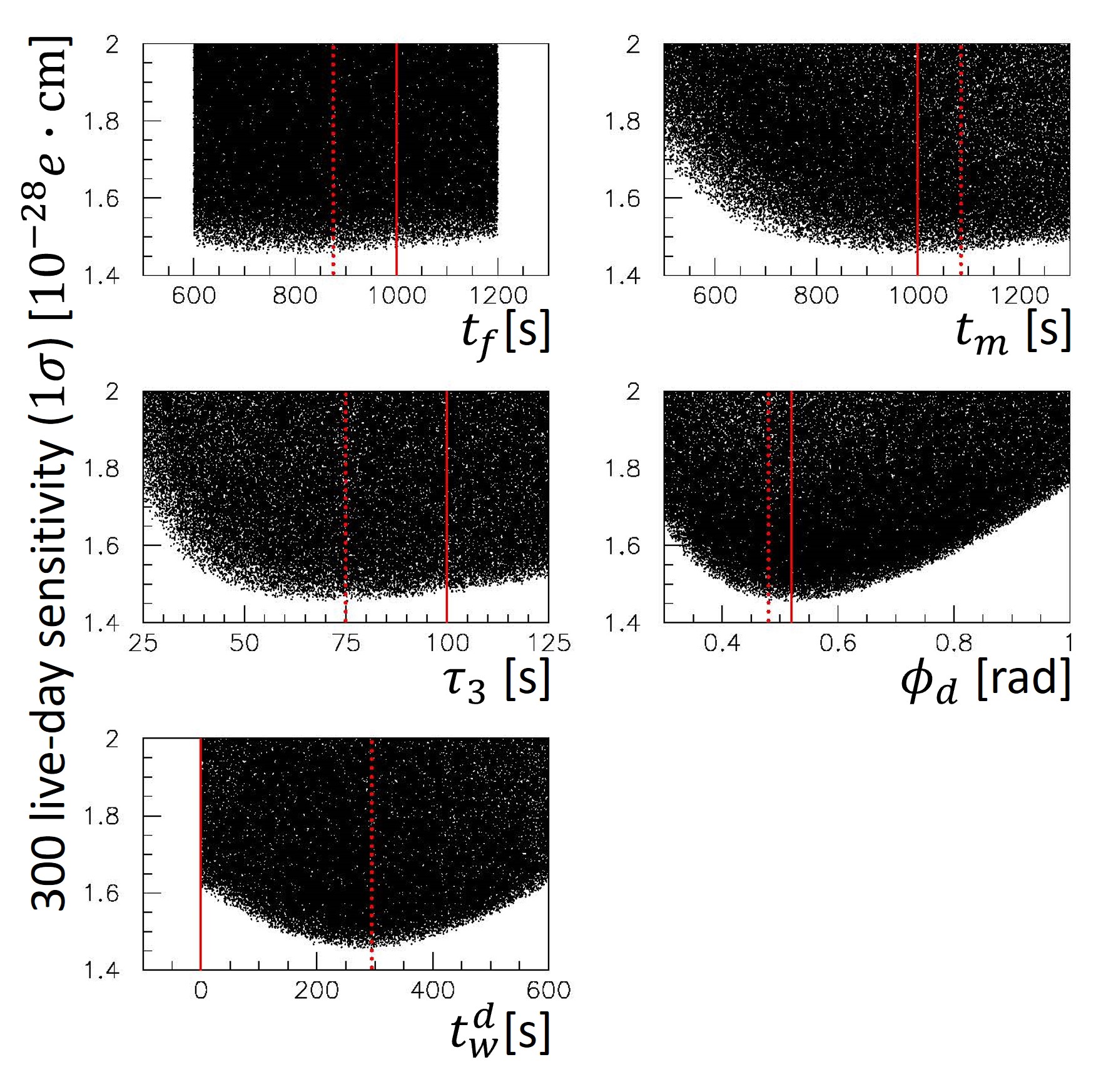}
\caption{nEDMSF sensitivity vs.~different operating parameter values, assuming performance 
parameters listed in Table~\ref{tab:pparam}.}
\label{fig:results}
\end{figure}

We next performed stochastic optimization of Equation~\ref{eq:final}. We started by setting
$\Delta t_0=1$\,s and $\Delta t_{\pm}=4$\,s%
\footnote{Hypothesisizing that dedication of 20\% of the live time to measuring backgrounds was a reasonable upper limit.} 
and allowing all other operating parameters to vary. We found an optimal sensitivity of 
$\sigma_d=1.45\times 10^{-28} e\cdot{\rm cm}$, a 10\% improvement over the result with the initial operating parameter values.
Results ($\sigma_{d_n}$~vs.~the different parameters) are shown in Figure~\ref{fig:results}. Vertical solid lines
show the initial operating parameter values.  Vertical dashed lines show the optimal operating parameter values,
which are also listed in the right-most column of Table~\ref{tab:oparam}. 
The only re-optimized parameter value that resulted in a significant sensitivity improvement is $t_w^d$. 
Note: optimal operating parameter values will depend on the actual performance parameter values.

We then removed the {\it in situ} background measurement by setting $\Delta t_0=0$ and
found the same optimal sensitivity. So, its inclusion has no deleterious effect on the 
experimental sensitivity and it provides significant benefits for later analyses.

The addition of the waiting period between the $\pi/2$ pulse and the start of $\phi_{3n}$
modulation improves the sensitivity because it minimizes UCN losses during the early 
part of the run, before the $d_n$-driven phase accumulation has had time to build up.
It seems plausible that gradually increasing $\phi_d$ over the course of a run would allow for further
optimization. To explore this we chose a hyperbolic tangent form for $\phi_d(t)$: 
\begin{equation}
  \phi_d(t_w^d<t<t_m) = \phi_d^{\leq}\left(\frac{\tanh((t-t_w^d-a)/b)+\tanh(a/b)}{\tanh((t_m-t_w^d-a)/b)+\tanh(a/b)}\right)
\end{equation}
\noindent Here, $b$ determines how sharply $\phi_d$ increases, $a$ determines the proximity of the 
sharpest increase to the end of the waiting period, the right-hand term in the numerator enforces the boundary condition 
$\phi_d(t_w^d)=0$ and the denominator enforces the boundary condition $\phi_d(t_m)=\phi_d^{\leq}$.
The step function change in $\phi_d$ at $t=t_w^d$ corresponds to $a=b=0$.
Both $a$ and $b$ were allowed to vary ($0<a<500, 1<b<1000$); no significant improvement was found.

We then simulated a full-statistics dataset, consisting of scintillation events and 
dressing state information ($\phi_d(t)$), and fit the data to Equation~\ref{eq:tofit}. 
$\phi_{3n0}$ was set to 0 in the simulation and fixed in the fit. The resulting 
uncertainty was $1.45\times10^{-28}e\cdot{\rm cm}$, in agreement with expectations. 

In a second simulated dataset, $\phi_{3n0}$ was allowed
to vary randomly ($\mu=0, \sigma=1$\,mrad) in 
each individual measurement. The overall uncertainty worsened 
to $1.6\times10^{-28}e\cdot{\rm cm}$. This follows from the strong correlation 
between $\phi_{3n0}$ and the extracted value of 
$d_n$ ($92\times10^{-28}e\cdot{\rm cm}/{\rm mrad}$). 
The resulting variation in $d_n$ is not insignificant compared to the $207\times10^{-28}e\cdot{\rm cm}$ 
statistical error on $d_n$ achievable in a single run, and the two add in quadrature to get the overall uncertainty.

\section{Conclusions}~\label{sec:conc}

The square-wave-modulated critical dressing method of the nEDMSF experiment can be
improved by including periodic background measurements (obtained by spending some time in each
modulation cycle with $\phi_d=0$). 
This greatly simplifies the fit function needed to extract $d_n$ and obviates the need 
for auxiliary measurements to measure the background's functional form. All this is 
achieved without sacrificing statistical sensitivity.

Optimized operating parameters were identified that improved the expected sensitivity 
by 10\%, to $1.45\times10^{-28}e\cdot{\rm cm}$. The improvement was primarily 
the result of adding a waiting period between the $\pi/2$ pulse that initiates $d_n$-driven 
$\phi_{3n}$ accumulation and the start of dressing modulation. 
This sensitivity was confirmed in a simulation of a full-statistics dataset.

At this level of sensitivity, 1\,mrad uncertainty in the initial $n$/$^3$He angle difference ($\phi_{3n0}$)is not insignificant.

\acknowledgments

The author gratefully acknowledges the support of the U.S. Department of Energy Office of Nuclear 
Physics through grant DE-AC05-00OR22725, and would like to thank
R.~Golub, C.~Swank and B.~Filippone for stimulating discussions,
and W.~Schreyer for carefully reading the manuscript.

\end{document}